\documentclass[english,aps,prstper,reprint,showpacs,titlepage,longbibliography,floatfix,nofootinbib,superscriptaddress]{revtex4-2}   

\usepackage[T1]{fontenc}	
\usepackage[latin9]{inputenc}	
\usepackage{geometry}		
\usepackage{graphicx}
\usepackage{tabularx}
\usepackage{subfig}
\usepackage{amssymb}
\usepackage[above,below]{placeins}	
\usepackage{times}

\usepackage{hyperref}  
\hypersetup{colorlinks=true,urlcolor=blue,citecolor=blue,linkcolor=blue}   
\usepackage{enumitem}          
\setlist{nosep}                 
\begin{document}

\begin{titlepage}

  \title{Same Activity, Divergent Impacts: Representing Paths Towards Physics Computational Literacy and Physics Identity with Conjecture Mapping-Based Narrative Analysis}

\author{Sarah McHale}
    \affiliation{Department of Engineering \& Physics, Providence College, 1 Cunningham Sq, Providence, RI, 02908} 
    \affiliation{School of Physics and Astronomy, University of Minnesota -- Twin Cities, 116 Church St. SE, Minneapolis, MN, 55455}
  \author{Tor Ole B. Odden}
    \affiliation{Department of Physics, Center for Computing in Science Education, University of Oslo, 0316, Oslo, Norway} 
  \author{Ken Heller}
    \affiliation{School of Physics and Astronomy, University of Minnesota -- Twin Cities, 116 Church St. SE, Minneapolis, MN, 55455} 


  \begin{abstract}
Integrating computation into physics teaching is a curricular move that, at present, has been predominately studied for its cognitive impacts. However, if this modality of instruction shifts how students engage with physics, we argue there is room for students to redefine what it means to do physics and how they perceive themselves relative to the field. To investigate this, we situate a comparative case study in the context of a computationally integrated physics course. We study two students' experiences with a multi-day activity to understand how and why they came to affectively divergent self-perceptions. We propose a modified use of conjecture mapping to visualize the production of affective outcomes and connect narrative analysis to activity design. Our analysis highlights how different interpretations of and engagement with activity design reflect students' epistemic framing of code, which, in turn, drives engagement with scaffolding in manners that shape self-perception.
  \end{abstract}
 
  \maketitle
\end{titlepage}

\section{Introduction}\label{Sec:introduction}
Both in physics and in STEM more generally, identity-based frameworks have frequently been used to study student success and persistence as a way to address demographic issues (e.g.,~\cite{hazari2010, stiles2018demystifying, aschbacher2010science, carlonejohnson, taheri2018structural, merolla2013stem}). Problematizing the metaphor of retention issues as a leaky pipeline, Stevens, et al.'s longitudinal ethnography of engineering students proposed a metaphor for the role that identification with the field plays in retention as that of a compass, guiding a student on their path through the field~\cite{stevens2008becoming}. This conceptualization illuminates the importance of the context in which students come to identify with their field of study.  

Many physicists have called for the teaching of computation in physics courses~\cite{caballero2020advancing, caballero2024computing,behringerengelhardt, behringer, oddencaballero, landau, chonackywinch}, and over the past few decades, progress had been made towards this educational goal~\cite{caballeromerner, fuller2006numerical,landau, chonackywinch, gavrin, picup, roos, paradigms, roundy2015look, caballero2014model, StThomas, cook2008computation}. However, studies into the ramifications of this change have predominately addressed cognitive rather than affective outcomes~\cite{araujo2008physics, CTframework, mackessy2021comparing, weatherford2013student, oddenzwickl, phillips2023physicality, lunk2012framework, kohlmyer2005student, oleynik2019scientific}. 
In reality, these affective and cognitive realms are deeply intertwined, as it has been well-established that physics students' beliefs can shape their learning and performance~\cite{haussler2000curricular, lishinski2017students, gupta2018exploring, elby2001helping, perkins2005correlating}. A similar phenomenon emerges in the scholarship that investigates integrating computing into a physics context, adding evidence that substantiates the claim that student beliefs shape student performance in intersections of computing and physics~\cite{bodin2012role, caballero2011evaluating}. 

When we integrate computation into the teaching of physics, we change the way physics is taught and thereby, we change the context in which students learn. If students orient themselves with the same compass of identification this new context, what happens to their self-perception? When integrating computation into student encounters with physics, there is potential to alter notions of what it means to be a physicist, as the ways students write, apply, and communicate code are added to expected physics skills. As such, what does it mean to develop a physics identity in a computationally integrated physics course, and how does that shape students' self-perception? Given that the cognitive impacts of learning computation alongside physics has been found to differ for students~\cite{bodin2012role}, we seek to understand where, how, and why the affective impacts differ for individual students.

Learning computation in a physics context may pose unique affective struggles, and as such, the affective realm is a meaningful site for investigation in and of itself. In interviews with students using computation in at least one upper-level college physics course, Lane and Headley studied student representations of computation relative to a professional community of practice and found partial alignment in reasons to use code, misalignment in areas of low confidence, and a community misunderstanding regarding common novice experiences~\cite{LaneHeadley}. In so doing, that study investigated students' perceptions of physics as a whole and did not probe how holding these beliefs shaped students' perceptions of themselves. Focusing on a different demographic, Hamerski, et al. studied experiences of high-school students in a computationally integrated physics class and clarified how computation can create additional frustration for students unfamiliar with computing by connecting those experiences to the theoretical constructs of self-efficacy, mindset, and self-concept~\cite{hamerski}. Instead, we are interested in what these encounters specifically do to shape students' self-perception, or identity, as physicists.

As part of a broader study on the affective impacts of a departmental effort to integrate computation throughout an entire physics major~\cite{mchale2024students}, we collected data from semi-structured interviews, mixed-method surveys, and ethnographic observations to understand the student experience in courses undergoing this curricular change process. In preliminary analysis, we found an emergent trend in our data: students expressed disparate self-perceptions in the wake of a deliberately designed multi-day activity. While we explored the difference between the professor's intent behind the design and the impact on a student in our companion paper~\cite{mchale2026evaluation}, in this paper, we aim to understand how and why two students came to divergent affective stances by investigating the following research question: \textit{How does engagement with the design of a multi-day activity in a computationally integrated physics course differently shape the development of students' physics computational literacy beliefs and physics identity?}

To address our research question, we 
use the method of conjecture mapping to represent narratives of two students' experiences with a multi-day activity. This method facilitates a visualization of the complexity of lived experiences that enables us to trace outcomes back to engagement with design through analysis of triangulated data sources.

In the form of a comparative case study~\cite{alma9975672698201701}, we examine the experiences of two students and, for each, first present a narrative of their experience. Then, we present the creation of a conjecture map to connect each student's narrative to the activity's design. 
The resulting conjecture maps serve as data visualization tools that aid in eliciting key themes from narrative analysis. Specifically, this method 
helps incorporate design into our narrative analysis in ways that further our understanding of what makes computing a differentially positive or negative experience for physics students.



\section{Theoretical Frameworks} \label{sec:theory}
In this case study, we characterize the impacts of integrating computation into the teaching of physics through an affective lens so that we can gain insight on how curricular design decisions foster or hinder the potential for students to have positive encounters with computation. Specifically, we consider affective outcomes of this modality of instruction by operationalizing students' self-perception through two theoretical frameworks: physics computational literacy and physics identity. 
\subsection{Physics Computational Literacy}
In a computationally integrated physics course, computation is complexly interwoven into the learning of physics. We therefore want to operationalize student perceptions of 
computing in a manner that can suitably analyze the rich affordances and difficulties of this curricular model. To do so, we  use the theoretical framework of physics computational literacy (PCL), which deliberately highlights the potential for computation to revolutionize learning in field-specific manners. 

PCL is a disciplinary example of diSessa's more general framework of computational literacy (CL)~\cite{disessa_2000}. To tap into the same transformative potential as the literacies of reading, writing, and mathematics, diSessa posited that a robust CL requires three pillars:
\begin{itemize}
  \item \textbf{Material CL:} familiarity and fluency with computing;
  \item \textbf{Cognitive CL:} applying computing to think differently; and
  \item \textbf{Social CL:} communicating with or about computing.
\end{itemize}
So, to understand what CL looks like in a physics context, Odden, Lockwood, and Caballero investigated how PCL could manifest, noting examples like~\cite{odden2019pcl}: 
\begin{itemize}
  \item \textbf{Material PCL:} familiarity with syntax, debugging code, algorithmic thinking, preferences for certain programming languages, beliefs about structure of code, etc.;
  \item \textbf{Cognitive PCL:} modeling practices, extracting physical insights from code, knowledge of suitability of computational methods, beliefs about how computation can be used in physics, etc.; and
  \item \textbf{Social PCL:} organizing code, commenting code, use of computational notebooks, beliefs about presentation of computational analyses.
\end{itemize}

Additionally, Odden, Lockwood, and Caballero highlighted that each pillar of PCL also spans a range of abstraction. While learning how to code in a physics context, students develop awareness of and comfort with PCL practices, knowledge, and beliefs. At present, the PCL literature mostly focuses on practices and knowledge~\cite{odden2019pcl, oddencaballero, oddenzwickl,fredly2026physics, cammarota2025social}. This leaves the affective component of PCL -- beliefs, defined as attitudes or value judgments -- lightly investigated, as the only study to directly probe such developments focuses on social PCL~\cite{nearhood_student_2025}. In this study, we focus our analysis on the development of beliefs across all three PCL pillars, centering students' reflections on their own PCL and how PCL is supported in the context of their computationally integrated physics course.

\subsection{Physics Identity}
We are also interested in understanding the role that encounters with computation play in the development of students' self-perception as it relates to identification with their field of study. To operationalize this in a discipline-specific manner, we consider students' physics identity, which is defined as the extent to which students see themselves as physicists or `physics people' \cite{hazari2010, hazari2020}.

This framework has its roots in the broader framework of science identity, developed through Carlone and Johnson's longitudinal study of successful women of color throughout their graduate and undergraduate education, into their early career in science~\cite{carlonejohnson}. The authors found recognition, performance, and competence to be meaningful factors that propelled these women towards success, but as the study focused on practicing scientists, they presupposed interest. When investigating the development of college students' physics identity, Hazari, et. al therefore introduced interest as another component, which their regression analysis found to be significant~\cite{hazari2010}. That same study also revealed that students did not distinguish between performance and competence as the members of Carlone and Johnson's study did, so those two components operated as a singular factor for students. 
A later study that accounted for students' progress towards degree found sense of belonging to be a quantitatively significant factor for the development of a physics identity, demonstrating a significant difference between first year and senior undergraduates~\cite{hazari2020}.

As this study centers on physics students at different points in their degrees, we consider the following as constructs of PI:
\begin{itemize}
  \item \textbf{Performance/Competence:} belief in one's ability to understand physics content and/or to perform physics tasks;
  \item \textbf{Recognition:} being recognized by others as a physics person / one's perception of how others view them in relation to physics;
  \item \textbf{Interest:} personal desire or curiosity to learn and/or understand more physics; \&
  \item \textbf{Sense of Belonging:} one's perception of fitting in or not feeling excluded in the/their physics community.
\end{itemize}

A wide range of qualitative, critical studies situated in a variety of contexts have investigated the perception, development, and performance of PI through the experiences of women of color, LGBTQ+ women, Black physicists, and neurodivergent physicists~\cite{ quichocho, hyaterCPI, mcdermott, quichocho2019does,hyater2018critical}. However, few investigations have focused on how students develop PI in a the context of a computationally integrated physics course. Such settings have the potential to position physics differently than the typical canon and further complicate what it means to be interested, competent, knowledgeable, and recognized in physics. Using a practice- and community-based model of identity, Bumler, et al. interviewed students in a computationally integrated introductory physics course and found that students developed new academic identities in such a context~\cite{bumler2019previous}. Their analysis concentrated on students' perceptions relative to coding more than physics, so there is more to be learned about how student encounters with computation shape students' perceptions of themselves as physicists. Instead, in this study, we aim to understand how the different ways in which students enact the design of a computational activity shapes the development of their physics identity.

\section{Methods}\label{Sec:methods}
\subsection{Data Sources}
As part of a broader study investigating the affective impacts of integrating computation into physics curricula, we collected ethnographic data in multiple computationally integrated physics courses, including the one that is the context of this case study. Our broader dataset consists of student and instructor interviews, document analysis, and course observations. 

In alignment with the overarching study's goals, we administered surveys to students in this class to gauge students' demographics, learning attitudes, self-efficacy, and beliefs about and experiences with computation~\cite{sosescp, class}. 
While the results of the surveys primarily served as a quantitative metric to monitor affective impacts for the broader study, we also used survey results to select interview participants whose responses spanned a range of self-efficacy, stances towards computation, demographics, and learning attitudes. Consequently, five students were invited to and consented to participate in a two-part interview series in which one interview occurred early in the semester and the other, towards the end. The first and third authors designed semi-structured interview protocols that prompted student reflection on their experience in the course as a whole through the lenses of PCL and PI. These interviews lasted approximately 1-1.5 hours, and interviews were recorded, transcribed, and deidentified for analysis. During transcription and analysis of the interviews, the first author used a researcher journal to reflexively understand and account for her role in the research process~\cite{ortlipp}. Upon preliminary analysis of the early-semester interviews, the first author also crafted tailored follow-up questions to address unique experiences. 

During the end-of-semester interviews, two students specifically mentioned a multi-day activity as a formative experience when asked about the development of their PCL and PI, with differing affective tones. In what follows, we present a comparative case study of the experiences of these two students during this multi-day activity. Both participants -- Bridget and Jake -- have been deidentified with student-selected pseudonyms and pronouns throughout this paper.

Since data was collected to study the course as a whole, the first author also interviewed the professor and TAs of the course, as their experiences were part of the focus of the overarching study. To do so, the first and third authors also designed similar two-part semi-structured interview protocols for the early-semester and end-of-semester interviews about instructional decisions and insight into implementation. As with the student interviews, these also lasted approximately 1-1.5 hours, and all interviews with consenting participants were recorded, transcribed, and deidentified for analysis. However, to keep student experiences at the heart of this case study, we primarily center the analysis of student interviews and use information from the instructor interviews to describe the setting of the case study.

To triangulate the interviews, the first author conducted classroom observations of the lab and lecture sections associated with this course and also attended Teaching Assistant (TA) meetings in which the professor and TAs planned and reflected on the operations of the course. Having served as a TA for this course in a prior semester, the first author engaged with students and instructors while observing and in so doing, built rapport with participants. Throughout observations, the first author recorded 46 pages of fieldnotes, and this raw data, as well as analytic memos, serves as foundational data for the study of this multi-day activity~\cite{emerson, tracy, SaldanaJohnny2016Tcmf}. These 33 hours of observations included the multi-day activity of this case study, as well as the TA meetings in which the activity was planned. Shortly after the conclusion of each observation session, the first author used a researcher journal to note connections between her observations and theory, paying deliberate attention to the impacts of her role on the research process~\cite{ortlipp}.

In addition, the first author also conducted document analysis of the course's Canvas page, including lecture slides, announcements, and submissions of student work. Analysis of these data sources provided an unobtrusive way to contextualize students' perceptions of the activity~\cite{bowen}.

\subsection{Conjecture mapping}
This study aims to characterize differing impacts of the design of a multi-day activity. To do so, we need an analytic method that can track the development of these outcomes back to the concrete modes of engagement with the activity design. The method of conjecture mapping, initially developed by Sandoval~\cite{sandoval}, is well-suited for such an investigation, especially given its roots in design-based research that also aims to improve educational practice.

Conjecture mapping begins with a conjecture, a theoretically-driven hypothesis that states how the design should operate. Since these high-level conjectures are too broad to govern design, these typically emerge from a preliminary analysis of the learning environment. 

The ways in which the learning environment reifies the conjecture are known as the \textit{Embodiment}, which Sandoval considered through \textit{Tools and Materials}, \textit{Task Structures}, \textit{Participant Structures}, and \textit{Discursive Practices}. Beyond considering the resources that students engage with, these embodiments also address the standards of the task, how participants are to engage, and the type and tenor of interactions. 

The design alone does not govern the development of outcomes, as participants need to engage with the design. Sandoval considered those through the category of \textit{Mediating Processes}, which includes interactions required to produce intended \textit{Outcomes}. Sandoval conceptualized mediating processes that emerge from design as \textit{Participant Artifacts} and \textit{Observable Interactions}, drawing attention to the need for triangulated data for such a method.

When connecting from an \textit{Embodiment} to a \textit{Mediating Process}, Sandoval considered the connection to be a \textit{Design Conjecture}, as it provides evidence that engagement with a specific component of a design surfaces in a specific manner. When connecting from a \textit{Mediating Process} to an \textit{Outcome}, Sandoval considered the connection to be a \textit{Theoretical Conjecture}, highlighting the commitment of design research to both the development of improved instruction and theory. For this to be the case, the outcomes need to be appropriately targeted and investigated.

In alignment with the cyclical nature of design-based research, Sandoval's original proposal of this method posed this as an iterative method, in which the same conjecture map would be iteratively revised as the intervention under study was modified from preliminary findings.  

\subsubsection{Modifications for Narrative Analysis}
In this study, we use conjecture mapping in a narrative form~\cite{oddenzwickl} so that we can compare conjecture maps that represent different students' experiences~\cite{wozniak}. Instead of iteratively revising the same conjecture map throughout the re-design of an activity, we create a conjecture map for each narrative that describes a different perspective on the same multi-day activity, and we analyze how and why the paths towards the outcomes differ. This enables us to meet the dual goals of iterative improvement and theory development while simultaneously considering and distinguishing different perspectives in a way that elucidates the structural ties to the development of these perspectives.

As such, we began the analytic process by writing narrative descriptions of each individual's experience with the multi-day activity. Our goal was to aggregate all of our data sources into stories that we would then use as data to analyze first-person accounts of this multi-day activity~\cite{tracy, merriam2019qualitative}. To do so, the first author began by combing through interview transcripts to isolate moments where participants explicitly spoke about what they did during the multi-day activity and how they felt about themselves as a result. Throughout this process, the first author recorded rationale for inclusion of data and noted emergent themes in analytic memos. The first author deductively coded the entirety of each individual's interview, in accordance with the theoretical frameworks as set forth in Section \ref{sec:theory}. Then, the third author independently coded snippets relevant to the multi-day activity while referencing a shared codebook. The authors iteratively discussed the meaning and assignment of codes until reaching consensus~\cite{tracy,SaldanaJohnny2016Tcmf}. As part of this conversation, the authors came to agreement on overarching affective narratives that encapsulated each individual's perspective on and experience with the activity. We did not use a quantitative method of inter-rater reliability, as it did not align with our analytic method of conjecture mapping, which relies on an iterative, interpretive process. However, throughout our analysis, we regularly referenced the triangulated data to ensure our methods and findings were rigorous and credible~\cite{tracy2010qualitative}.

In a similar manner, the first author analyzed fieldnotes from the relevant observations to triangulate students' reflections on their own engagement with what the first author witnessed. The first author then combined data from fieldnotes and interviews to create a chronological narrative description of each individual's perspective on and experience with this multi-day activity. Then, the first author conducted a document analysis of each participant's submitted work for the associated homework assignments. Because the purpose of this case study is to understand the affective impacts rather than cognitive impacts, these documents served to provide context rather than shift the underlying narrative; our analysis prioritized students' reflections on their work more than our interpretation of their work.

\begin{figure*}
  \includegraphics[width=\textwidth]{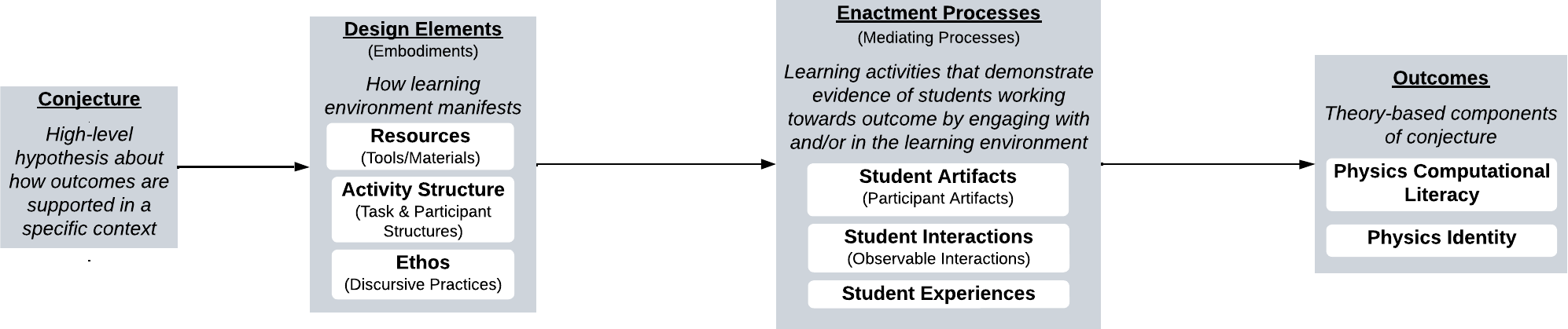}
  \caption[Components of a conjecture map, as used to visualize triangulated narratives.]{Components of a conjecture map. When we use a modified term, Sandoval's original term is included in parentheses~\cite{sandoval}.}
  \label{fig-ConjectureMap2}
\end{figure*}

We then used our modification of Sandoval's original construction of conjecture maps to conduct a narrative analysis, as detailed in Figure~\ref{fig-ConjectureMap2}. For narrative analysis, we use embodiments differently because we are not considering the perspective of the professor who designed the activity and concretized a conjecture in the learning environment. Instead, when we consider embodiments, we consider how the student interpreted the design of the activity, which may well diverge from the instructor's intent as explored in our companion paper~\cite{mchale2026evaluation}. So, to address this, we mirror Boelens, et. al's use of conjecture mapping and recategorize embodiments as \textit{Design Elements}~\cite{boelens2020conjecture}. To more concretely tie our analysis to a single activity, we reconceptualized Sandoval's categories as \textit{Resources}, \textit{Activity Structure}, and \textit{Ethos}. These components still cover support mechanisms, design considerations, and discursive practices, as Sandoval's original categories did.

Similarly, our use of mediating processes differs, as there is no expected engagement, only a reporting of the engagement that students used. So, in a manner similar to Boelens, et al., we rename this category to be \textit{Enactment Processes}, highlighting the ways in which students enact design by through the work they produce and the ways they interact in this learning environment~\cite{boelens2020conjecture}. This still necessitated further refinement of what we considered to be \textit{Enactment Processes}. Inspired by Kim and Saplan, we rename Sandoval's category of Participant Artifacts as \textit{Student Artifacts} for clarity of analysis because we analyze student experiences. Further, in studying affect, we found it important to broaden the ways students enact a design, so we followed the work of other scholars and created the category of \textit{Student Experiences} to address engagement that did not solely rely on artifacts or interactions~\cite{boelens2020conjecture,kim2024making}. 

The way in which we use \textit{Outcomes} does not meaningfully diverge from Sandoval's original proposal, as they are theory-based components of our conjecture. We connect to an outcome when a participant's interview data was deductively coded as a specific component of PCL or PI. Although, as our conjecture maps aim to communicate the development of these affective outcomes, the meaning of our linkages somewhat differs from Sandoval's original use. On our conjecture maps (as detailed in Fig.~
\ref{fig-ConjectureMap2}), the first set of connections (Sandoval's design conjectures) represent what participants did, whereas the second set of connections (Sandoval's theoretical conjectures) represent how participants felt about themselves as a result of their engagement. We also use these connections to represent the valence, or the underlying tone categorized as either positive or negative~\cite{russell1999bipolarity}. We acknowledge that this analytic decision turns complex emotions into a binary, but we do so to allow for nuance in representing the development of affect. Students can experience a debilitating struggle while engaging in their learning environment, and we consider this under the umbrella of negative valence. This draws a distinction between similar experiences perceived as a generative struggle, which we would consider with a positive valence. As such, we represent positive affect with solid lines and negative affect with dashed lines on both participants' conjecture maps. 
In essence, these trajectories aim to communicate each individual's perspective on the multi-day activity. 

The first author created these connections while eliciting the design elements and enactment processes as they emerged from the narratives. The narrative, rather than discrete data sources, then served as the foundation from which to connect a design element to an outcome through an enactment process and served to characterize the affective connection as either positive or negative. While drafting initial conjecture maps, the first author constructed original subcategories of design elements and enactment processes for each participant, clearly tying each connection on the conjecture map to specific data sources. Eventually, categories deemed to be redundant within an individual's narrative were collapsed through reflexive analysis in the researcher journal and communication with the research team, but the consolidation of components did not propagate across different participants' conjecture maps. The first author created the initial conjecture maps independent of each other and recorded rationale for all decisions in analytic memos. Naturally, there were elements of design and modes of engagement that both students mentioned and by working chronologically, the first author had already established appropriate terminology. In those circumstances, the production of one map may have informed the other.

To ensure a rigorous and robust creation of the conjecture maps, the second and third authors read the first author's narratives and compared them with the respective conjecture map.  Determining what to call each enactment process was a consensus-based practice, in which all authors discussed the narrative, the suitability of each enactment process, and the valence underlying the affective connection. 

With rigorous and robust conjecture maps in hand, the first author then compared the two student maps to each other. Again, these comparisons were shared with the second and third authors, who discussed their thematic takeaways from comparing both narratives and conjecture maps.

In the analysis that follows, we do not present conjecture maps that represent these students' entire experience with and perception of the multi-day activity, as those maps are highly complex. Instead, to demonstrate a meaningful use of the method, we present a notable theme that emerged through the first author's use of analytic memos and analysis discussions within the research team: these students conceptualized and engaged with code differently. As the second and third authors read the accompanying narratives, the research team consolidated narratives to the most salient components for the theme and only included connections on the conjecture maps relevant to those components. 

\section{Comparative Case Study}\label{Sec:case_study}
To elucidate the ways in which the affective impacts can differ in individual students' encounters with computation in physics, we situate our comparative case study in a computationally integrated modern physics lab course. 
Throughout a multi-day activity, two students, deidentified as Bridget and Jake, developed differing affective stances towards their PI and PCL. We first set the scene in Section~\ref{context} by presenting an overarching description of the course context and the more specific setting in which the multi-day activity occurred. Then in Sections~\ref{Bridget_narrative} and~\ref{Jake_narrative}, we present narrative descriptions of the two students' perspectives. In Sections~\ref{Bridget_map} and~\ref{Jake_map}, we summarize the narratives while explicating the creation of each conjecture map. We then bring high-level summaries of these conjecture maps into conversation with each other in Sections~\ref{compare} to tie our narrative analysis to activity design.

\subsection{Setting \& Context} \label{context}
More broadly, this case study occurs within an institutional setting that is beginning to integrate computation into the physics major. In specific, we focus on the first course in which the department has integrated computation: a modern physics laboratory course. This sophomore-level class is a requirement for both physics and astrophysics majors and is designed to be the first course that prospective majors take beyond the introductory sequence. As the introductory sequence of courses is excluded from this computational change effort, this is the first course in which students encounter computation. 

Each week, 39 students met for a single 50-minute lecture in which the professor discussed the statistics behind analysis methods students used in their weekly lab. Students split up into 3 sections of 13 students for a weekly 3-hour lab with a graduate Teaching Assistant (TA). In the first lab of the semester, students completed introductory activities to familiarize themselves with MATLAB, the computing platform of the course. Then, they conducted a common lab, in which all students performed the same experiment and used MATLAB to conduct a statistical analysis of their results. This introduction to analysis laid the foundation that linear fits would be the central analysis method of the course.

Over the course of the 14-week semester, students moved from using built-in functions to writing their own linear fit functions. Similarly, the results students were expected to report advanced in sophistication, from fit parameters -- slope and intercept -- with uncertainty to fit parameters characterized by $\chi_{r}^2$. This progression in scaffolding was reinforced in both homework assignments and in lab, and when students were expected to use a new skill, they regularly encountered it first in homework. 

Students often needed to linearize their data before conducting a linear fit, as many experiments explored phenomena described by power laws. In the final few weeks of the course, the professor, deidentified as Professor Evans, introduced a multi-day activity that extended students beyond the linear fits they had used all semester and introduced the concept of non-linear fitting. 

This multi-day activity began with a homework assignment that asked students to refer to an earlier homework in which students determined the fit parameters for a linearized power law relationship and found the corresponding uncertainties. Students could use their previous findings to calculate $\chi_{r}^2$, but correct numerical values were also provided for the fit parameters and corresponding uncertainties. Importantly, this new approach did not require students to linearize their data, as they already knew the appropriate fit parameters. The assignment asked students to hold one of their fit parameters constant while coding a \texttt{for} loop to vary the other within the established range 
to produce a plot of the resulting 2-dimensional $\chi_{r}^2$ surface. This same homework assignment then asked students to use nested \texttt{for} loops to vary both fit parameters and plot the resulting 3-dimensional $\chi_{r}^2$ surface and explain the graph.

This was the first time in the semester that students were asked to use \texttt{for} loops, so Professor Evans spent time in lecture on the day that she assigned this first associated homework to direct students to specific references to learn about \texttt{for} loops. Professor Evans knew this activity was a stretch for students, as she had used it in previous semesters in which she taught the course. This semester, she provided students with a pre-coded grid search function that would find the fit parameters that minimized $\chi_{r}^2$ and that students could use for the second associated homework. For a second homework assignment during the next week, students would conduct a nonlinear fit by finding the fit parameters, rather than using known fit parameters, as they did in the first associated homework assignment. This would not require students to write the more complicated code themselves, and it also provided a clear structure through which the multi-day activity built on itself. This redesign of the multi-day activity included an in-class workshop during which students worked to understand the provided grid search function. This workshop was the midpoint between the two associated homework assignments, giving students time to review their solutions from the first assignment and use the provided code to complete the second assignment. 

This provided grid search function was designed and delegated during TA meetings. In a TA meeting the day before the in-class workshop, a TA deidentified as Luca mentioned that he was still modifying the previous year's activity and writing the grid search function Professor Evans wanted students to use; he did not have the function finalized at the time of the meeting. Professor Evans and Luca discussed some slight issues that surfaced in his current version. The solutions for the first associated homework assignment were not completed at the time that Luca wrote his solutions for the in-class activity, so Luca did not reference them when writing his grid search function. Luca continued to edit his code throughout the meeting and sent it to Professor Evans later that evening. During this meeting, Professor Evans framed the multi-day activity as "a module on complicated stuff" and highlighted that for that reason, she would not be grading students' code annotations for accuracy. Luca said he thought his code might be simple enough for students to figure out, but Professor Evans stressed that it needed to be doable. She admitted that she thought students would not get to the stage of producing the desired plot during the in-class workshop, which was a component of the second associated homework.

The following lecture section was the in-class workshop in which students were given the 2-dimensional grid search function that found the fit parameters that minimized $\chi_{r}^2$. The classroom was arranged as it was for every lecture session -- rows of tables arranged parallel to the projector screen -- and most students sat in the same seats that they had occupied throughout the semester. To ensure students had a solid understanding of the provided code, Professor Evans printed paper copies of this function and asked students to both bracket each \texttt{for} loop and write pseudocode explaining what each line of the provided function did. She specifically highlighted that even though the code was commented, students should describe it in their own words. Professor Evans also made sure to clarify that the code was also on the course Canvas page. 
Before releasing students to work on the activity, she presented slides that introduced nonlinear fitting and described how a grid search algorithmically minimizes $\chi_{r}^2$. Her final slide included instructions for the in-class activity and second associated homework assignment, providing guidance about the code annotation and encouraging students to determine what changes the function would need to fit to the error function rather than a power law. These instructions remained displayed for the duration of the class. 

Even though it was not written on the slides, Professor Evans verbally clarified that students would receive half of the points for the subsequent homework assignment for participating in the in-class workshop. This point distribution was also addressed in the document explaining the second associated homework assignment, which was posted the day before the in-class workshop.
She then released students to get to work, and the classroom remained silent, except for the rustling of papers. Professor Evans then reminded students that "at first it might be quiet as you read things through, but feel free to ask questions of your peers," and continued to remind students that they could ask questions both to their peers and to her throughout the entire workshop. 

Early in the workshop, a student approached the first author, who was observing the course, and asked for clarification on what appeared to be an unused variable that closely mirrored that of another declared variable. The first author read the code, communicated the irrelevance of the declaration, and asked Professor Evans for confirmation. However, Professor Evans clarified that she did not write the code so thus did not know, but she did know that the code Luca sent worked. So, the first author contacted Luca, who replied to the first author 15 minutes later and confirmed that the variable was no longer used. The first author shared this with Professor Evans who then announced to the class that this variable was unused and could be ignored. At the end of class, Professor Evans wrapped up the workshop by stating that it was "an example of using someone else's code" and reminded students that they were able to change it if they wanted.

The tasks that students were asked to complete as part of the in-class workshop were explicitly included in the second associated homework assignment, which was the final homework assigned in the course. In addition to the components addressed during the in-class workshop, the final assignment also asked students to create a 3-dimensional plot of the $\chi_{r}^2$ surface, with the path taken by the grid search overlaid.

\subsection{Bridget's Perspective}
\subsubsection{Bridget's Narrative} \label{Bridget_narrative}
Bridget is a sophomore who intends to major in both physics and astrophysics. In a pre-course survey, she reported that she expected to put in 2-5 hours per week outside of class and to earn an A in this class. There, she also reported that she was enrolled in 9-12 credits this semester and worked 11-20 hours per week outside of class. Bridget's survey responses also indicated that prior to this semester, she had conducted research at her university, but in interviews, she expressed that she felt that even though her research used code, it wasn't the primary skill she developed.

Instead, Bridget's formative computational experience was when she  
initially learned how to code in a computer science (CS) course. In her early semester interview, Bridget noted that even though that course used Python, she believed it equipped her with broad, transferable skills, which she was able to apply in this computationally integrated physics class: 
\begin{quote}
    If I hadn't taken a coding class prior to [using] MATLAB, then it would be very new because coding is, yes, it's different for every language, but once you understand how to assign a variable, that's really all you need to do and learn how to graph things. So there's some things that are specific to MATLAB that I need to search up, but for the most part, just the inherent like how a coding program works was helpful because I had already taken a class.
\end{quote}
Bridget felt that this CS class was "very strict on sharing code," but she understood why: 
\begin{quote}
    You need to do the code on your own because everyone codes different. So they would crack down on if you shared code with other people. So it was more like, `Hey, did you figure out the logic behind this problem?' every once in a while rather than getting together and working on it.
\end{quote}

On the first homework assignment associated with this multi-day activity, Bridget found it "very exasperating" to struggle with "some error that [she] wasn't able to troubleshoot." So, 
she met up with a classmate to discuss the code she had written for this homework -- an approach she rarely used in this course. She chose to meet up with her classmate in person out of concerns for academic honesty: 
\begin{quote}
    If you're sending [your code electronically], there's that possibility where they can just copy paste or copy the exact same thing, but when you're showing it [in person], it's more like they're understanding the process of how you did it, `cause no one's going to memorize in the two minutes you showed them your thinking process.
    \end{quote}
Even though this physics class did not have rules for collaborating on code, Bridget extended her CS course's policy to this class and centered her collaborative conversations around the ideas behind her code rather than specific syntax. This then precluded her from resolving issues with her code:
\begin{quote}
   I reviewed my thinking process and code with someone else and we had the same thought process... I had implemented it concept wise, but I wasn't able to debug it fully. And then because of that, the next homework assignment didn't really click as easily as well.  
\end{quote}
Our document analysis revealed that the issue Bridget was trying to debug was not in fact a syntax issue -- it was a conceptual issue. While Bridget thought something was wrong with her code, instead, something was wrong with her linearization. To successfully complete the first associated assignment, Bridget should have log-transformed her data, but she passed raw data to her linear fit function and therefore got unrealistic fit parameters. While it was not necessary to resolve this issue to successfully complete the remainder of the multi-day activity, this early hurdle set Bridget up to internalize a negative self-perception throughout the activity.

Frustrated, Bridget compared herself to her peers, as she knew that she had classmates who were more comfortable with MATLAB and did not struggle with the assignment to the extent that she did. Even though she reported spending copious amounts of time on the assignment, she felt badly about herself for not achieving success:
\begin{quote}
   It made me feel as if I didn't have enough motivation. So I would try and I'm like, I'm not getting anywhere, but I felt like if I put more time and effort and I had that motivation to do so, I could have improved.
\end{quote}

Bridget felt that choosing to collaborate mitigated this affective damage, because she also mentioned that talking through her code with her classmate helped her move from viewing her own difficulties as a personal weakness to seeing her and her classmate in a shared struggle. This reframing increased her sense of belonging through a form of solidarity that she described as, "if we're going down, we'll go down together." 

Nevertheless, Bridget's collaboration did not resolve the challenges she faced in the first associated homework assignment. So, she turned to an approach she had used throughout the course -- reviewing solutions. However, as explored in our companion paper~\cite{mchale2026evaluation}, Bridget did not view this activity as coherent with the course as a whole. Bridget's perceived lack of coherence also impacted her epistemic framing of the use of code, as she believed that the centrality of code used for lab analysis was reinforced in the course's overall grading scheme. Bridget justified her general decision to focus more on lab analysis than anything else in the class "because it's worth the amount of points it is." Lab analyses composed 40\% of students' course grades, whereas homework composed 10\%. This imparted the message to Bridget that "it's good to work on the homework, but if you don't 100 percent get it right, they post the solutions. And so it's good to review those and they all build up to the [lab] analysis." By regularly being able to review solutions to improve the code she wrote for homework for future use in lab, Bridget came to believe that it was more important to learn from her mistakes than to submit perfect code. 

Previously, this conceptualization benefited her, as she could recycle code she had previously written and apply it in a new context. Bridget reported doing this "all the time" and found it "super easy to recreate" what she had successfully implemented in a different context. However, 
the multi-day activity 
frustrated Bridget because she felt it "was more simply MATLAB knowledge based" and did not focus on the application, as she knew she would not use the code she would write in lab analyses. So, activating the same framing during the multi-day activity did not prove as fruitful for Bridget as it typically was throughout the course, even though the activity was designed to build on itself and offered the opportunity for her to recycle code as she progressed through the stages of the activity. In fact, this compounded for Bridget when she encountered cognitive and affective barriers early in the activity.

After facing challenges with material PCL in the first associated homework assignment, Bridget hoped that reviewing solutions would help her learn the syntax she thought was the root of her struggle, stating: 
\begin{quote}
  I just couldn't quite grasp it, so it was more of I'm just going to turn in what I have. So it didn't quite feel that good. It didn't make me feel confident... 
It was more like I'll turn in what I have and once the homework answers are posted, I can go through that and see how they're able to do it. 
\end{quote}

However, the solutions to the first associated homework assignment were not posted until the day of the in-class workshop. Bridget therefore carried forward both confusion and frustration with material PCL from her initial encounter with the multi-day activity into the in-class workshop, as she did not have confidence in the foundational syntax on which she was was to build. All in all, Bridget's inability to come to an initial understanding of her own code made each stage of the multi-day activity feel "equally exasperating" as she then needed to understand code that was provided to her.

At the start of the in-class workshop, Bridget was silent, listening to Professor Evans' contextualization of the task at hand. She sat alone but proximal to other students, and the first author observed that Bridget spent the majority of class individually bracketing each \texttt{for} loop in the script. 
Then, Bridget's level of engagement changed when one of her peers noticed the error in the script -- the unused variable. When her classmates nearby started talking about the unused variable, Bridget said "I'm still doing all the brackets," focusing on coming to her own understanding of the code. Shortly thereafter, Bridget altogether disengaged from the course activity and began to engage in off-topic conversation with her surrounding peers. Despite the resolution that arose 15 minutes later, Bridget's social conversations extended beyond the period of uncertainty and continued through the end of the class period. She never moved to annotating pseudocode that explained how the provided grid search worked. However, during her end-of-semester interview, Bridget reflected that "having it written really helped [her] understand it."

Although Bridget had gained confidence both with the general act of programming in her CS course and with the ways she used MATLAB in this class, she reported that she had struggled to transfer that confidence to 
this in-class workshop because of the language-specific difficulties she faced. Despite her familiarity with components of code like nested \texttt{for} loops, Bridget found it difficult to understand code written by someone else that relied on language-specific functions:
\begin{quote}
It felt like a huge jump to me simply because like I mentioned, there were just parts in MATLAB or MATLAB code that I wasn't familiar with. So I've been doing well with the coding for MATLAB, but it's typically been more generic and simple code like straightforward code that I would use. And this involved a lot more complicated code. So I understand \texttt{for} loops and if else statements et cetera, but using specific MATLAB functions was a bit more difficult to understand.
\end{quote}
With the exception of one homework assignment that provided code to produce a figure, students had never been asked to use code written by someone else. Further, students had never been asked to write pseudocode, and even though Bridget reported that having a paper print-out of the provided function helped her understand it, she still left the in-class workshop without a full understanding of the provided function. So, to gain intuition on its operation, Bridget attempted to run the function on her laptop after class. Although, the second associated assignment asked students to do more than use the code; they were asked to modify the provided code. Having not achieved a baseline understanding, Bridget struggled to determine how to modify the code to such an extent that she never submitted the second associated assignment: 
\begin{quote}
  They provided the code on Canvas, so I just copy-pasted it and ran it through after. But it was a bit confusing as to what I was supposed to do to make the code work. 
\end{quote}


\subsubsection{Bridget's Conjecture Map} \label{Bridget_map}
\begin{figure*}
  \includegraphics[width=\textwidth]{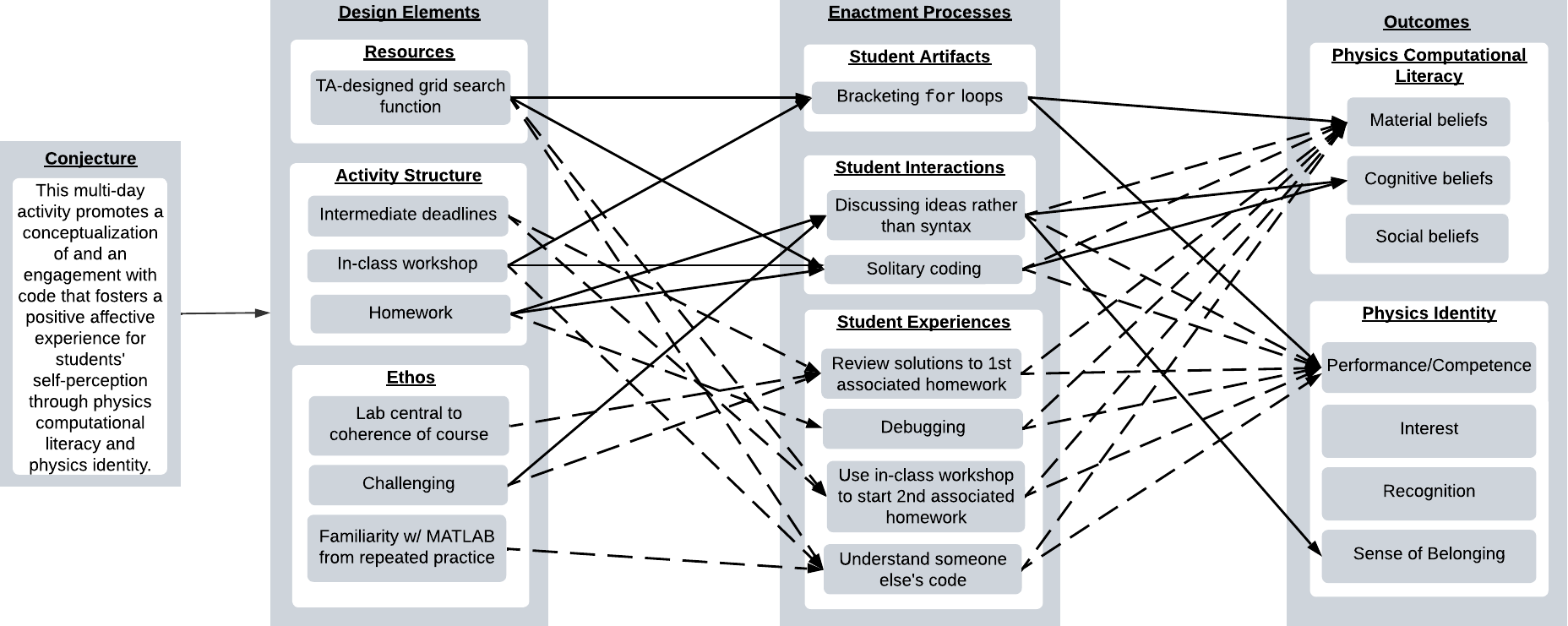}
  \caption[Bridget's conjecture map, representing her conceptualization of and engagement with code during the activity under study]{Conjecture map representing Bridget's perspective on the multi-day activity, specifically depicting her conceptualization of and engagement with code. Connections with positive valence are shown as solid arrows, and connections with negative valence are shown as dashed arrows.}
  \label{fig-Bridget-narr}
\end{figure*}

We begin our narrative analysis by proposing a conjecture that relates to the theme underlying this case study: \textit{This multi-day activity promotes a conceptualization of and an engagement with code that fosters a positive affective experience for students' self-perception through physics computational literacy and physics identity.} Our representation of the activity's impact on Bridget's PCL and PI is given in Figure~\ref{fig-Bridget-narr}. We populate the outcomes with the constructs of PCL and PI, our two theoretical frameworks for affect. We then use Bridget's narrative to determine the enactment processes used and what affective role those serve in the usage of design elements and the production of affective outcomes. In the analysis that follows, we move temporal chunk by temporal chunk to begin to trace Bridget's narrative on a conjecture map. 

When working on the first homework associated with the multi-day activity, Bridget faced issues with her code that she was unable to debug. As part of this struggle, Bridget negatively evaluated her own material PCL, which we consider to be a negative material PCL belief and a negative evaluation of her own performance/competence. We consider homework to be the relevant design element on which she struggled. So, we connect from the design element \textit{Homework} to the enactment process of \textit{Debugging} and to the outcomes of \textit{Material beliefs} and \textit{Performance/competence}, all with negative affect. 

Bridget initially approached the first associated homework assignment independently, so she when she faced difficulties that she attributed to material PCL early on, she internalized them as personal shortcomings. However, because she felt she understood the concepts, this also meant that she attributed her mistaken confidence in cognitive PCL to herself. Because Bridget felt good about her decision to work alone, embodying her collaborative practices from her CS class, Bridget's conjecture map connects positively from the design element of \textit{Homework} to the enactment process of \textit{Solitary coding}. 

Since Bridget felt she struggled more with the syntax than the implementation of ideas while working alone, she developed frustration with material PCL and her own performance/competence while still maintaining confidence in her cognitive PCL. Even though Bridget's perception of her difficulty contrasted with our analysis of her work, we foreground her own reflections. We therefore connect from the enactment process of \textit{Solitary coding} to the outcomes of \textit{Cognitive beliefs} with positive affect and to \textit{Material beliefs}  and \textit{Performance/competence} with negative affect. 

To address these challenges, Bridget met up with a classmate outside of class. She applied the same collaborative method from her CS course, which emphasized cognitive PCL but prevented her from successfully debugging her difficulties with material PCL, as she couldn't display the surface plot. So, we connect from  the design elements of \textit{Homework} and \textit{Challenging} to the enactment process of \textit{Discuss ideas rather than syntax} with positive affect. To represent that Bridget faced the same issues when collaborating with her peer that she also faced independently, we therefore connect from the enactment process of \textit{Discuss ideas rather than syntax} to the outcomes of \textit{Cognitive beliefs} with positive affect and to \textit{Material beliefs}  and \textit{Performance/competence} with negative affect.

In addition, choosing to collaborate with a classmate also provided Bridget with a form of solidarity that displaced her insecurity in not knowing. Bridget moved from viewing her own difficulties as a personal weakness to a shared struggle, and in so doing, created a counterweight to her negative performance/competence appraisal by connecting positively to physics identity through an increased sense of belonging. This is represented on Bridget's conjecture map as a positive connection from the enactment process of \textit{Discuss ideas rather than syntax} to her physics identity through the outcome of \textit{Sense of belonging}. 

Because of the limited extent to which homework counted towards her final grade in this physics course, Bridget believed it was more important to learn from the homework solutions than to submit perfect homework. Typically, Bridget reviewed homework solutions to learn from her mistakes, as that benefited her in a more central component of the course, the lab. So, Bridget established a pattern of engagement and wanted to turn to this same support mechanism to resolve her confusion in the early stages of the multi-day activity. Unfortunately, Bridget could not do so because the solutions to the first associated homework assignment were not posted until the day of the in-class workshop. So, we represent this unproductive struggle on Bridget's conjecture map with a negative connection from the design elements of \textit{Intermediate deadlines}, \textit{Lab central to coherence of course}, and \textit{Challenging} to the enactment process of \textit{Review solutions to 1st associated homework}. The lack of resolution led Bridget to inherit confusion and frustration with material PCL across the multi-day activity, so this same enactment process connects negatively to the outcomes of \textit{Material beliefs} and \textit{Performance/competence}.

During the in-class workshop, Bridget worked alone when bracketing the provided grid search function. So, the design elements of \textit{In-class workshop} and \textit{TA-designed grid search function} both connect positively to \textit{Bracketing \texttt{for} loops} and \textit{Solitary coding}. Even though Bridget never reached a full understanding of the provided function, she felt that bracketing the paper copy helped her understand the code more than if she had engaged with it digitally. This surfaces on Bridget's conjecture map with positive connections from \textit{Bracketing \texttt{for} loops} to \textit{Material beliefs} and \textit{Performance/competence}.

Engagement with the TA-designed grid search function was a critical component of the in-class workshop, but Bridget struggled to understand the provided code because it was written differently than how she would have and did not have the same preparatory scaffolding for understanding code written by others as existed for other computational tasks. So, we connect from the design elements of \textit{Familiarity w/ MATLAB from repeated practice}, \textit{In-class workshop}, and \textit{TA-designed grid search function} to the enactment process of \textit{Understand someone else's code} and to the outcomes of \textit{Material beliefs} and \textit{Performance/competence}, all with negative affect. 

This difficulty also impacted Bridget's engagement with the activity, as she solely engaged in off-topic conversation after encountering the component of the code written differently than she knew. As this then shaped the extent to which she struggled to start the second associated homework assignment, we connect from the design elements of \textit{TA-designed grid search function} and \textit{Intermediate deadlines} to the enactment process of \textit{Use in-class workshop to start 2nd associated homework} with negative affect.

After the in-class workshop, Bridget tried to engage with the provided function by running the code but still was unable to understand how it worked, which she needed to do to be able to modify the code for the second associated homework assignment. To represent this, we negatively connect from the enactment process of \textit{Use in-class workshop to start 2nd associated homework} to the outcomes of \textit{Material beliefs} and \textit{Performance/competence}.

\subsection{Jake's Perspective}
\subsubsection{Jake's Narrative} \label{Jake_narrative}
Jake is a junior Astrophysics major. In a pre-course survey, he reported that he expected to put in 6-10 hours per week outside of class and to earn a B in this class. Jake was enrolled in over 16 credits this semester and worked 1-10 hours per week outside of class. While Jake had never conducted research before, he had numerous encounters with computation prior to this course. 

Jake first learned to code in eighth grade, when he began to tinker with programming his calculator. Consequently, he became comfortable with self-guided, informal avenues for computational support, which helped him learn by seeing how other people wrote code:
\begin{quote}
    I remember I used some just abhorrent programs, just line after line, no breaks, no organization, anything happening. But over time I learned... I watched a lot of YouTube videos to pick up [another coding language], so I got a lot of like, oh, this is how they're organizing it, and that makes a lot of sense. I like the way they're doing that. Now I'm going to do that. Kind of just picking up the skills of how to organize code through that.
\end{quote}
Since then, Jake has taken multiple CS courses and a university-level math class that used MATLAB. These experiences helped him develop a broad familiarity with many coding languages. He now identifies coding as one of his passions and actively chooses to code in his free time.

Although Jake felt he was already computationally literate before the course, he felt even more confident in his material skills after the course because of the way he could leverage familiarity with another language on the job market.
More broadly, Jake conceptualized this as a course on "data handling," which he liked because it aligned with how he perceived the utility of computing in his subfield of interest. 
This was the first physics class that Jake had taken that used code, and it showed him the "front end, the face of what is presented by physics to everybody else." This made him understand what a career in physics could be like, and Jake was excited to see coding as an activity of a physicist. He felt he gained "a skill of application because writing code for a physics situation is going to feel different than writing code for some other sort of situation."  

For the entirety of the multi-day activity, Jake worked alone. At the start of the in-class workshop, Jake hunched over his printout of the provided code while annotating and kept his head exceptionally close to the paper. Occasionally, students sitting nearby would ask him a question, and Jake would answer while remaining bent over his code printout, continuing both to bracket and 
write pseudocode explaining the functionality of the code in his own words. The first author could see Jake's code printout and saw that he had thickly annotated it. Though Jake's physicality and participation reflected a level of focused commitment, he did not exhibit signs of frustration.

Once it appeared that Jake had finished annotating the code, he spent the remainder of class time helping his peers. 
Then, after drawing the first author's attention to the unused variable, a classmate sitting near Jake uttered to herself, "I'm really confused." In response, Jake immediately turned around and gestured at his laptop, where he had the code pulled up. When this classmate continued to ask questions, Jake later used his pencil to point at specific lines on his classmate's print-out of the provided code. This classmate was the one who pointed out the inconsistency with the unused variable, and Jake explained to her and her group the utility of the built-in function that it resembled, trying to address her confusion. The inconsistency did not derail Jake's productivity as it did for other students, and he continued explaining the code to his peers for the remainder of the class period. Other students who sat near Jake but not immediately next to him came over to sit by him and asked questions. Jake's conversations often focused on broader concepts like step sizes and 2-dimensional spaces, but he repeatedly gestured to specific lines of code to tie his conceptual explanations to concrete syntax. 

Even though the inconsistency of the unused variable did not derail Jake's ability to understand the function, he 
was not satisfied with the provided grid search and decided to improve it outside of class: 
\begin{quote}
I disliked the code so much that I decided to rewrite it, so I had a good understanding of it. So I was able to offer a lot of help. That felt really good. I also learned a lot about the concept that way by just going through and just digging into the code side of it and just figuring out how the code works helped figure out how the physics is working. Kind of convenient!
\end{quote} 

Of particular importance, Jake believed that coding helped him understand the physics through his process of organizing conceptual knowledge into specific subcomponents when rewriting the grid search function. With this activity in specific, Jake felt that his comfort with code made learning the physics more accessible: 
\begin{quote}
    [Using code] definitely helped me work through physics problems and really get a feel of them in a way that fits in my brain better... Like when I was rewriting that code, I was reading through the textbook a lot about this whole method of how we're doing the non-linear fitting, and it wasn't sticking very well, but when I was rewriting it through the code and reading the other code, trying to figure out what the heck it's doing and rewrite it in my own way, I mean, it told me the same thing the physics textbook was trying to tell me, but I was receiving that information in a way that put it together differently in my head and in a way that I felt more comfortable and familiar and just lined up with me better. I feel like it gave me the information in a more familiar form.
\end{quote}

Because Jake had frequently communicated with Professor Evans during the semester, he decided to share his rewrite with her. Professor Evans then asked if she could share it with the class and use it in future implementations of the activity, which was a profound moment of recognition for Jake:
\begin{quote}
    It was really cool to have that kind of more opportunities budded out from doing that... I think it kind of represents how I'm being seen in terms of that.... It's extremely rewarding. It feels like an achievement to have gotten myself to the point where I can do that kind of thing and kind of reach an extra level.
\end{quote}

Even though he did not submit the first associated assignment, Jake's deep and meaningful participation in the in-class workshop led him to receive a perfect score on the second associated assignment.

This multi-day activity stuck out to Jake as a positive moment in a demanding semester. Jake reported that this semester had been challenging because of a heavy credit load and a long daily commute. Across his courses, he was struggling to succeed, but he felt more supported by Professor Evans than his other professors. At this point in the semester, Jake felt that there was not much he could do to improve his standing in his other courses, but he felt that he could improve his grade in this course, which he considered possible, in part, because his other classes had begun to taper off to prepare for finals.

Jake had a long commute to campus, which infringed upon his ability to collaborate with his classmates, who regularly met up in-person. Even though Jake preferred to meet up with people in the physics and astrophysics major lounge, his schedule often prevented him from doing so. Instead, he often turned to an online platform, Discord, that many of his classmates used for both academic and social support. The asynchronous and digital nature of Discord enabled Jake to both give and receive help in "quite an equal" balance. In other physics classes, he was "usually the one asking questions," but in this class, Jake felt the balance between giving and receiving help was different, as he was "able to offer a lot of help with the MATLAB stuff." Throughout the course, Jake used the same mode of engagement that he used with the provided grid search function when he worked with other classmates, reflecting that "most of the time I'm trying to understand someone's code, I just translate it into what I'm familiar with." 

In particular, Jake felt especially good about the amount of help he could offer with this multi-day activity. After the in-class workshop, Jake's peers engaged with him both through direct messages and through the broader group messaging board: 
\begin{quote}
    With that last assignment with the code thing, I had several people [direct message] me and trying to talk to me in the broader Discord... It is super good to feel helpful and, I guess it's kind of weird to say, valuable. Maybe it's more, I don't know, you feel necessary.... It definitely makes me feel like a part of the [physics] community. I have a place in the community and this is that place.
\end{quote}

While Jake felt recognized by his peers as being a good coder, a skill this class associated with physics, this raised a contrast for Jake when he realized his classmates didn't ask him questions about physics. However, because he felt like a "generally confused physics person," Jake did not take issue with this:
\begin{quote}
    It's nice to feel like they probably see me as somebody who knows what they're doing in terms of code. I wouldn't blame them and I don't expect them to think that about what I'm doing in terms of knowing what I'm doing in terms of physics.
\end{quote}


\subsubsection{Jake's Conjecture Map} \label{Jake_map}
\begin{figure*}
  \includegraphics[width=\textwidth]{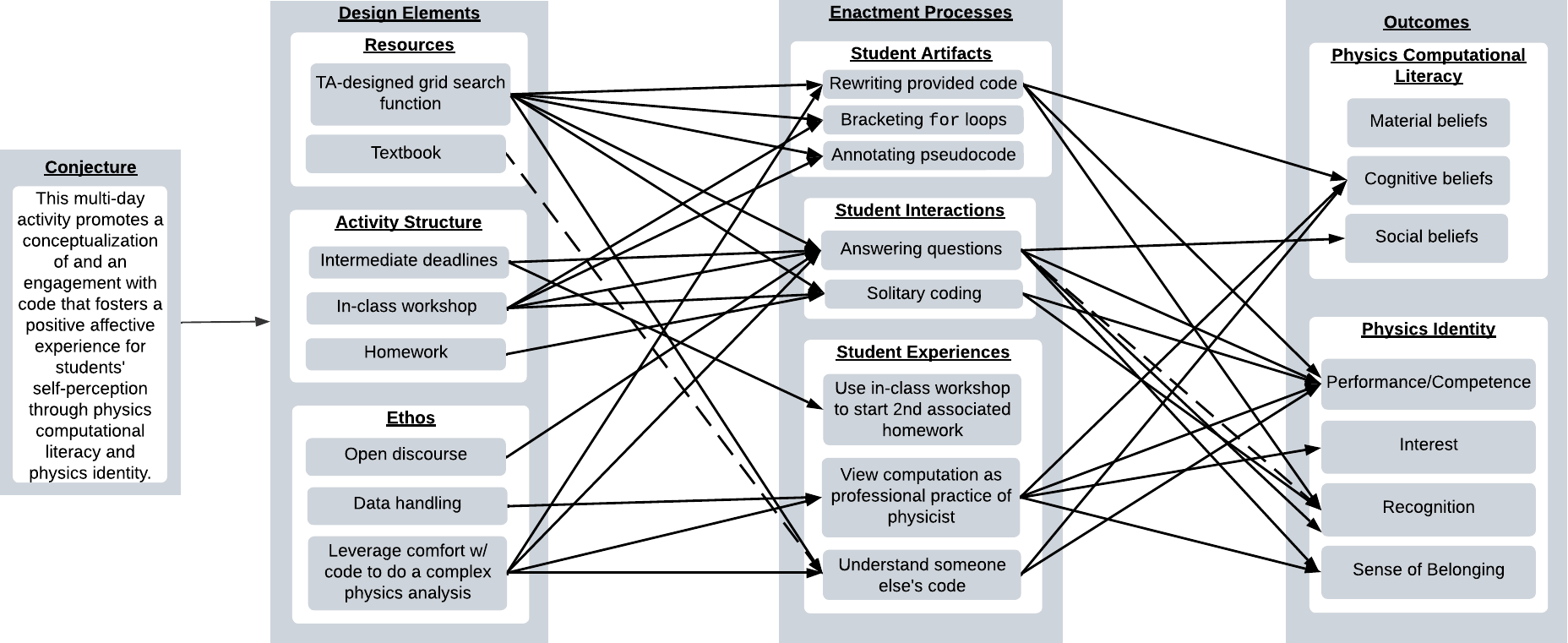}
  \caption[Jake's conjecture map, representing his conceptualization of and engagement with code during the activity under study]{Conjecture map representing Jake's perspective on the multi-day activity, specifically depicting his conceptualization of and engagement with code. Connections with positive valence are shown as solid arrows, and connections with negative valence are shown as dashed arrows.}
  \label{fig-Jake}
\end{figure*}

We represent the activity's impact on Jake's PCL and PI in Figure~\ref{fig-Jake}. As with Fig.~\ref{fig-Bridget-narr}, we explore the same conjecture: \textit{This multi-day activity promotes a conceptualization of and an engagement with code that fosters a positive affective experience for students' self-perception through physics computational literacy and physics identity.} We also populate the outcomes with the constructs of PCL and PI, our two theoretical frameworks for affect. We then use Jake's narrative to determine the design elements that he engaged with, as well as  the enactment processes used, to characterize what affective role those served in the the production of affective outcomes. In what follows, we move through Jake's narrative chronologically to create his conjecture map.

Jake approached this activity with the understanding that computational tasks served the role of data handling, which he appreciated because of how it helped him prepare for a career. This was especially important for Jake because he already felt confident and comfortable coding before this class, so gaining physics-specific skills met a need. In other words, Jake was already confident in his material PCL and this offered him an opportunity to gain cognitive PCL, which he was eager to learn. So, we positively connect from the design element of \textit{Data handling} to the enactment process of \textit{View computation as professional practice of physicist} and to the outcomes of \textit{Cognitive beliefs}, \textit{Performance/competence}, and \textit{Interest}. 

During the multi-day activity, the first author observed Jake working alone while bracketing and annotating the provided function. As the observations showed Jake working thorugh complexity with confidence, this surfaces on the conjecture map in positive connections from the design elements of \textit{TA-designed grid search function} and \textit{In-class workshop} to the enactment processes of \textit{Bracketing \texttt{for} loops}, \textit{Annotating pseudocode}, and \textit{Solitary coding}. 

Jake also used the in-class workshop to share his computational expertise and help struggling classmates. So, we positively connect from the design elements of \textit{In-class workshop}, \textit{Open discourse}, and \textit{TA-designed grid search function} to the enactment process of \textit{Answering questions}.

Frustrated with the solutions provided during the in-class workshop, Jake rewrote the code in a way he could understand better and said that this, in turn, helped him understand the physics. Positive connections from the design elements of \textit{TA-designed grid search function} and \textit{Leverage comfort w/ code to do a complex physics analysis} to the enactment process of \textit{Rewriting provided code} to the outcome of \textit{Cognitive beliefs} reflect this experience on Jake's conjecture map.

The affective boost in cognitive PCL beliefs was important to Jake because he struggled to understand the physics concepts underlying the code when he turned to the textbook as a source of support, as shown by a negative connection from the design element of \textit{Textbook} to the enactment process of \textit{Understand someone else's code}. 

Understanding the provided code by rewriting it helped Jake feel that he understood the underlying physics better by enabling him to translate the information into "a more familiar form." We represent this by positively connecting from the design element of \textit{Leverage comfort w/ code to do a complex physics analysis} to the enactment processes of \textit{Understand someone else's code} and \textit{Rewriting code}, both of which positively connect to the outcomes of \textit{Cognitive beliefs} and \textit{Performance/competence}. As Jake referenced the provided function throughout this process and felt a highly productive form of stress in disliking the provided function, we connect from the design element of \textit{TA-designed grid search function} to the enactment process of \textit{Understand someone else's code} to the outcome of \textit{Cognitive beliefs}, all with positive affect.

Jake then shared his rewrite with Professor Evans. Because he created this new version of the code alone and interpreted the professor's reaction as an acknowledgment of a skill, we positively connect from the enactment processes of \textit{Rewriting provided code} and \textit{Solitary coding} to the outcomes of \textit{Performance/competence} and \textit{Recognition}.

Jake did not submit the first associated homework assignment. Nevertheless, he submitted the second homework assignment, which he started during the in-class workshop, and received a perfect score. So, on Jake's conjecture map, the design element of \textit{Intermediate deadlines} connects positively to the enactment process of \textit{Use in-class workshop to start 2nd associated homework}. Additionally, Jake worked alone and only discussed homework with other students when he was explaining what he had done, after completing it. So, we also positively connect from the design element of \textit{Homework} to the enactment process of \textit{Solitary coding}.

The temporal scaffolding also offered the opportunity for Jake's classmates to continue to ask him for help after the in-class workshop. His continued connections with his classmates demonstrated how he carried over the underlying ethos of the in-class workshop, that students were supposed to ask questions of, and in turn help, each other. Jake thoroughly enjoyed discussing code with his classmates and sharing his expertise. This experience helped him to feel like his classmates saw him as a valuable part of the community for this skillset he had to share. To reflect this, we connect from the design elements of \textit{Open discourse}, \textit{Intermediate deadlines}, and \textit{Leverage comfort w/ code to do a complex physics analysis} to the enactment process of \textit{Answering questions} and to the outcomes of \textit{Performance/competence}, \textit{Recognition}, \textit{Social beliefs}, \textit{Sense of belonging}, all with positive affect. 

These engagements helped Jake feel like he found his place in the field of physics, now that computation was an associated skill. We represent this by connecting from the design element of \textit{Leverage comfort w/ code to do a complex physics analysis} to the enactment process of \textit{View computation as professional practice of physicist} to the outcome of \textit{Sense of belonging}.

Still, Jake knew that being recognized for his coding skills wasn't the same as being recognized for his physics skills, which he did not feel confident about. So, we also connect from the enactment process of \textit{Answering questions} to the outcome of \textit{Recognition} with negative affect.

\subsection{Cross-case Comparisons} \label{compare}
Comparing Bridget and Jake's conjecture maps enables us to visualize the similarities and differences in their experiences across their interpretation of design elements, uptake of enactment processes, and clustering and valence of the connections between components of the conjecture maps.


\subsubsection*{Different Design Elements: Bridget and Jake interpreted the ethos of the activity differently.}
In a testament to the clarity of the structural implementation of the multi-day activity, Bridget and Jake used many of the same design elements. They both engaged with the TA-designed grid search function during the in-class workshop and noted the importance of intermediate deadlines for homework. Differences between the two students' interpretation of the design surfaced in the category of ethos -- how they interpreted the spirit of the activity. 

Bridget viewed lab as central to the coherence of the overall course, which is not surprising as this is labeled as a lab course. She carried this framing into the multi-day activity, even though the activity would not be directly applied in lab. This stands in contrast to Jake, who viewed the course's emphasis to be on "data handling," which he viewed as extending beyond the classroom and tightly connected to how he conceptualized the multi-day activity and what he wanted to do as an astrophysicist. Jake's understanding of the activity was therefore more in alignment with the task at hand. This divide set Bridget up to inherit negative affect, and Jake, positive.

Broadly, Bridget found the multi-day activity to be challenging, which motivated her to collaborate with a classmate. However, being unable to resolve her confusion led her to hope she could review solutions and learn from her mistakes in a solitary manner. As these were not posted with enough lead time before the in-class workshop, 
Bridget struggled to benefit from reviewing solutions because her confusion had already catalyzed a negative self-perception that left a fog of negative affect throughout the rest of her engagement. This was not the case for Jake, who understood the code and was therefore able to capitalize on his professor's request for students to work together. He enacted a climate that prioritized open discourse among students by answering the questions of his peers, and Jake felt good about himself in doing so.

Similarly, Bridget attempted to apply her familiarity with MATLAB from repeated practice throughout the course, even though the ways in which this multi-day activity used code differed with respect both to difficulty level and to form of scaffolding. So, Bridget struggled to understand someone else's code, which was a necessary component to engage with the scaffolding of the provided grid search function. Her struggle to do so led her to develop a negative PCL belief about her own material literacy because of what she considered to be shortcomings in her understanding, which we categorize as performance/competence. With confidence from numerous prior computational experiences outside of this course, Jake, on the other hand, was able to bypass the computational difficulties of this activity and see them as an opportunity to leverage his comfort with code to do a complex physics analysis. In fact, his computational multilinguality was so robust that it helped him overcome frustrations when he could not understand how the textbook was trying to explain the concept of non-linear fitting. Jake's ability to understand someone else's code allowed him to focus on the underlying physics, and in so doing, he developed confidence in his performance/competence due to a bolstered cognitive PCL belief. 

\subsubsection*{Different Enactment Processes: Bridget and Jake's prior computational experiences shaped their engagement with the design.}
Bridget and Jake also engaged with and produced artifacts differently in this activity. While both students bracketed the provided code print-out, only Bridget highlighted the role that this specific mode of engagement played in her understanding of the syntax and her resulting confidence in material PCL. Nevertheless, that source of support proved insufficient in the face of additional sources of negative affect that overwhelmed Bridget throughout the activity, and as a result, her self-perception was largely negative. Jake, on the other hand, did not mention engagement with the printed code as a meaningful source of engagement and identity-building, even though observation showed that he was fully engaged with the activity. However, his engagement with the printed code was more holistic, as he both annotated and rewrote the code. It was the rewriting of the code, rather than the engagement that was a designed component of the activity, that was more impactful for Jake's PCL and PI. While Jake used a technique he was familiar with to understand the provided code, Bridget did not have that wealth of experience to draw upon and found it challenging to understand code written differently than she would have written. 

In the realm of student interactions, both students prioritized independent work, but when they moved to engage with their peers, they did so differently. Because Jake deeply understood the code, he had confidence to engage with his peers with limited intersubjectivity and offer help rather than ask for it. 
Bridget, alternatively, worked with a peer who had a similar level of understanding, so her connections from solitary coding mirrored those from discussing ideas rather than syntax -- a preconceived notion she carried over from the collaborative policy in her CS course. Whereas Bridget developed solidarity with a peer, Jake experienced a profound moment of validation for his individual strengths.

These differences also surface in the category of student experiences. Bridget struggled to start the second associated homework during the in-class workshop because she was unable to debug the first assignment. So, she carried over confusion, but moreover, frustration with herself. Having to build off of this foundation quickly during the in-class workshop meant that this meant that the scaffolding, in the form of the TA-designed grid search function and intermediate deadlines, was insufficient to address Bridget's difficulties and to provide the affective support she needed.

So, when facing challenges early in the multi-day activity, Bridget wanted to review solutions to 
understand where she went wrong. Even though she believed that everyone wrote code differently, she wanted to reference with a source of authority. On other assignments, doing so had helped her achieve success in the lab component of the course. But, on the multi-day activity, when she later struggled to understand the provided grid search function that was written by a TA, Bridget disengaged and instead chose to engage with her peers through off-topic conversation because of a small inconsistency in the code that directly impacted the extent to which she could understand the task at hand. The primary difficulty for Bridget was not what the scaffolding existed to support -- nested \texttt{for} loops. Bridget had encountered those in her CS class and knew how to write them. Instead, her primary challenge was understanding someone else's code that was written differently than how she had been asked to write code this semester. Specifically, the built-in MATLAB functions confused Bridget because she had never seen them before and they were not explained in the code's comments. The new disruption to this tight pattern of coherent engagement highlights how strongly Bridget relied on transferability between uses of code.

Jake, on the other hand, had practice with computational translation. His computational training surfaced in the support mechanisms he used when faced with confusing code. Rather than attributing fault to his own understanding, as Bridget did, Jake rewrote the code, as he often did when coding for fun. His initial informal, self-taught training helped him feel comfortable seeing other people's code and developing an understanding of how what he wrote related to someone else's syntax. Jake was not afraid to drastically modify code to make it more understandable to him. 

However, three of Jake's enactment processes, \textit{Bracketing \texttt{for} loops}, and \textit{Annotating pseudocode}, and \textit{Use in-class workshop to start 2nd associated homework}, served as dead ends, not leading to outcomes. We attribute this to the fact that his computational strengths allowed him to circumvent some of the scaffolding, so these enactment processes did not shape his self-perception. Even though Jake did not complete the first homework assignment, he did not need it to build the skills necessary to successfully execute the rest of the multi-day activity. Jake did not note the importance of starting the multi-day activity in the workshop by bracketing \texttt{for} loops and annotating pseudocode for his affective development because, we claim, he was not reliant on the scaffolding to experience success. 

\subsubsection*{Different Clustering and Valence of Connections: The impacts on Bridget and Jake's PCL and PI were imbalanced.}
Bridget regularly evaluated her material PCL and her performance/competence simultaneously. Inspecting her conjecture map reveals that she had numerous sources of negative affect but only one source of positive affect towards those two constructs. The fact that Bridget positively evaluated her cognitive PCL becomes complicated by our document analysis of her submission, as we see the error that she carried forwards, which might have been the source of most, if not all, of her negative affect. What she thought was an issue with material PCL was in fact an issue with cognitive PCL, but Bridget did not view it that way. Even though she felt that using code in the context of a lab was the most important way code was framed in this class, she did not evaluate herself on those metrics. She felt it was necessary to have competence in material PCL and blamed her code even when her conceptual understanding of how to conduct physics analyses was the problem. However, this affective damage was mitigated when Bridget collaborated with a classmate on the first homework and developed a sense of belonging in knowing that she was not the only one who struggled.

In comparison to Bridget's conjecture map, connections to Jake's outcomes look markedly different. He only had one negative affective outcome, knowing that the recognition he was receiving in this activity was atypical for him. Jake became aware that if he was a source of help for computing, he was not a source of help for physics. However, this did not dominate his overall affective experience, as he had numerous sources of positive affect. Furthermore, Jake's affective outcomes were well-balanced across PCL and PI. The only component which he did not experience a shift in self-perception was material PCL. We pose that this aligns with Jake's claim at the start of the semester that he was already comfortable and confident with the act of coding in many languages. Jake already had a computational multilinguality that equipped him with the skills necessary to grow in the other components he was less well-versed in. 

\section{Discussion}\label{Sec:discussion}
In this case study, we used conjecture mapping as a tool to visualize narrative analysis. Doing so elucidated how two students came to different affective outcomes through divergent interpretations of the design's ethos, enactment of the activity, and patterns of connection between map elements. Across each of these contrasts, we see a common driving force: Bridget and Jake 
held different epistemic framings regarding the role of code. These led the students to both interpret and engage with the activity design differently, resulting in divergent affective impacts on their self-perception.

\subsubsection*{Epistemic framings can be inherited from prior computational exposure.}
More broadly, framing refers to one's conceptualization of an event at hand~\cite{goffman1974frame, tannen1993framing, tannen1987interactive}. Specifically, epistemic framing addresses how students perceive what knowledge they will generate and how, considering the context in which they are expected to do so~\cite{hutchison2010attending, shaffer2006epistemic}. With these definitions in mind, our case study revealed how Bridget and Jake entered this multi-day activity with pre-existing epistemic framings of code and activated them in a manner parallel to the activation of cognitive resources~\cite{lunk2012framework,hammer2005resources}. 

Before this computationally integrated physics course, Bridget's only experience with code was in a formal CS course. Bridget's understanding of why coding was important in this class was informed by her understanding of the course's grading mechanisms. By embracing this, she internalized what was important from her professor's structural messaging, despite activity-specific guidance about collaboration. 
However, despite the lack of course-specific guidance around collaboration, Bridget activated her epistemic framing of code from her prior computational training in a CS class, where she came to believe that she should discuss ideas behind her code rather than the syntax itself. This framing surfaced in the first homework assignment associated with this multi-day activity, when she attempted to collaborate to address the struggles she attributed to material PCL. The scaffolding, both in the form of intermediate deadlines and provided code, did not address Bridget's conceptual difficulties when she inherited an issue from the early stages of the activity. Bridget had established a pattern of engagement that relied on the coherence of the lab-centric course, and as such, she came to both write and understand code that reflected that coherence. In addition, when facing a conceptual issue she could not debug, she found the activity to be challenging and felt it was "equally exasperating" throughout. She wanted to turn to the solutions to come to an understanding of where she went wrong. While a qualitative case study of a computationally integrated high school physics course attributed a desire to see provided answers to a fixed mindset~\cite{hamerski}, we take a slightly different angle. We pose that all of these threads embody a hierarchical epistemological framing of code, in which Bridget framed knowledge construction as guided by an external authority. This mirrors a framing cluster seen in traditional physics instruction, where physics students invoke authority as an epistemological resource (e.g.~\cite{hammer2005resources, bing2009analyzing}). 

In contrast, Jake began his computational training in playful, exploratory programming environments. He felt comfortable in many languages, so on this activity, he was able to leverage that in an atmosphere of open discourse. While Jake engaged with the scaffolding provided during the in-class workshop by annotating the print-out and writing pseudocode, he reflected more positively on his use of a familiar technique. He both rewrote the provided grid search function to understand it and translated his classmates' code when helping them. In so doing, Jake engaged with code provided by the professor in the same way that he engaged with his peers' code, reflecting a non-hierarchical epistemological stance towards the construction of code, as he did not frame knowledge as more valid when it came from an authority. 

However, since Jake did not ask questions of his peers or his professor, his lack of reciprocal engagement with others suggests a limited intersubjective orientation. He positioned himself as an independent sense-maker rather than a co-constructor of knowledge. Bridget shared a similar solitary epistemic framing, but when she experienced failure, she utilized her intersubjective orientation to discuss ideas rather than syntax.


This difference encapsulates divergent interpretations of collaborative norms in the realm of computing. Bridget did not want to discuss the syntax underlying her code, and Jake felt comfortable doing so to help his peers, which bifurcated their affective experiences when Jake student experienced success and Bridget didn't. In an interview study with students enrolled in upper-level physics courses with computational homework assignments, Lane and Headley saw a similar phenomenon as we saw in Bridget's experience when students in their study wanted to develop code from scratch and feared plagarism when consulting external sources, despite the fact that expert programmers regularly search for sample code~\cite{LaneHeadley}. Our analysis highlighted the potential for lack of messaging around collaborative norms to serve as a catalyst for compounding cognitive inequities into affective inequities, as Jake, who had more computational experience, could leverage that in ways Bridget was unable to leverage her limited prior experience and feared collaborating for purposes of academic dishonesty.

\subsubsection*{Scaffolding, both cognitive and affective, shapes epistemic agency in ways that shape how students attribute their own success, or lack thereof, to themselves.}
By coming into the activity with divergent epistemic framings about the construction of code, Bridget and Jake differently valued the computational knowledge they could create and the knowledge they would gain by doing so, which shaped the ways in which they attributed their success, or lack thereof, to their own doing. Bridget understood code to be something that she could eventually perfect from by comparing her work to that of an authority and internalized the struggle as a lack of motivation when she faced issues she could not resolve. Jake instead treated others' code as highly malleable and valued his own implementation. This addresses the idea of epistemic agency, as Bridget and Jake regularly evaluated both the steps they took to build knowledge and the resulting outcomes of those processes~\cite{stroupe2014examining, ko2019opening, miller2018addressing, sundstrom2023instructing}. In particular, our case study highlights the affective realm of epistemic agency and indicates that students simultaneously make judgments about their self-perception relative to the field of physics when enacting different amounts of authority in the production of knowledge.

Research on nontraditional introductory physics labs has highlighted how the structure of the labs and the role of the instructor impact epistemic agency, and in addition, has suggested that lab instructions can also shape both students' epistemic framing and instructor's actions~\cite{sundstrom2023instructing}. We consider the scaffolding of the provided grid search function to be a continuation of the latter thread. In this case study, we found that interactions with scaffolding can activate different epistemic frames that propel students towards divergent affective outcomes. In so doing, our case study expands upon research that explicated how prior computational exposure does not ensure a seamless transition to computation in a physics context~\cite{hamerski,LaneHeadley}. Connecting these affective experiences to students' epistemic framing helps extend other studies on epistemic framing and epistemic agency in computational encounters with physics to the impact on students' self-perception~\cite{bodin2012mapping, odden2023using}.

Similarly, in a study of a computationally integrated engineering course, Vieira, et. al found that despite the benefits of early scaffolding, students struggled to understand later projects in the course, which they attributed to the increasing difficulty of the projects over the course of the semester~\cite{vieira2015exploring}. They recommended that "for the projects where the disciplinary content is overwhelming for students, the scaffolding should be focused on the specific subjects instead of providing too much computational support" and called for further research on the cognitive ramifications of different modes of scaffolding. While that study focused on cognitive impacts of scaffolding, our case study investigated the development of students' self-perception and found that scaffolding can also produce divergent affective outcomes. In a parallel of how Vieira, et al. found that students who engaged with greater amounts of scaffolding may have been less engaged in higher-level thinking, we find that Bridget, who was in a position to benefit more from the scaffolding's assistance, felt less epistemic agency when engaging with it. By not needing the scaffolding, Jake could attribute his success to himself and avoid the trap Vieira et al. discuss of using scaffolding to complete tasks but struggling to interpret results.

In many ways, the provided grid search function aligns with a common scaffolding technique for teaching computation in physics: minimally working programs (MWP). A MWP is a functional but incomplete portion of code that students are expected to modify to gain physical insight~\cite{mwp}. Often, these programs include a visual component. By providing students with a function that they were asked to modify on the following homework assignment, Professor Evans effectively gave students an abstract MWP, without the visual component. Especially at the introductory level, MWPs are a popular way to scaffold computational skills in a physics class, but this approach has only been studied for its cognitive impacts~\cite{oleynik2019scientific,lunk2012framework,weatherford2013student,Irving_2017, mwp}. Nonetheless, our affective investigation found similar salient phenomena. Weatherford, et al. found that a feature as small as a variable name in a MWP could distract a student deidentified as Norma~\cite{weatherford2013student}, and the unused variable served a similar role for Bridget, even though its inclusion was accidental. Since Weatherford, et al. investigated the cognitive impacts of the MWPs, they saw Norma resolve her confusion when she engaged with the visual output of the code. In our case study, we did not see Bridget resolve her negative affective stance, even after the professor's clarification of the cognitive resolution. Jake did not feel encumbered by this inconsistency in the code and instead occupied a position as a knowledge-holder in explaining the operation of the grid search function to his peers, which was a catalyst for his positive self-perception. With these contrasting experiences in mind, our case study highlighted that MWPs have the potential to create affective impacts on students' self-perception through connections to epistemic agency.

\subsubsection*{Computationally integrated physics activities can nuance our understanding of how students view themselves in relation to the field.}
The resulting impacts on Bridget's self-perception were reinforced across multiple forms of engagement, and she repeatedly evaluated her material PCL at the same time as her performance/competence. The co-occurrence of these connections on her conjecture map also highlight the parallels of her underlying affective tone in developing these outcomes. The role of competence in identity formation has surfaced in studies of other computational encounters in physics, but it surfaced through belonging~\cite{LaneHeadley}. In our case study, Bridget's experience reveals how even if the development of code is decentered relative to the use of code, that does not mean students evaluate themselves on that metric. To use code, Bridget needed to feel confident in her understanding of the syntax, which was especially important given that she would not discuss syntax when collaborating. A tight connection between material PCL and performance/competence also underlies Jake's experience; he had a deep understanding of code and a multilingual comfort that not only helped him succeed but also enabled him to be recognized for his coding skills.

In non-computational contexts, the literature on PI expresses a general coherence around the importance of recognition in the formation of a PI~\cite{hazari2020, kalender2019gendered, sundstrom2025bias,  lock2013physics, kalender2019female}. One component Wang and Hazari found key for recognition was that of attainable success~\cite{wangHSPI}. Their study highlighted how, in addition to cognitive scaffolding, affective scaffolding becomes especially important in the realm of dealing with and overcoming failure -- a phenomenon Wood, Bruner, and Ross refer to as frustration control~\cite{wood1976role}. In our case study, unsuccessful engagement with the activity's scaffolding led Bridget to work with a peer. While she made progress towards frustration control in doing so, she did not experience success and instead developed a sense of solidarity that justified her confusion. This approach helped Bridget develop a positive sense of belonging despite her negative material PCL beliefs, but it did not set her up to attain success in later stages of the activity. 

Jake, on the other hand, experienced a level of success in this activity that was especially impactful because of explicit recognition from his professor and peers when his work was shown as exemplar code. Nair and Sawtelle studied a similar episode 
in an introductory physics course for non-majors, highlighting the powerful reinforcement of such recognition when a student can bring in more advanced code from another class~\cite{nair2019operationalizing}. Analyzing that course from an ecological systems perspective, they noted how explicit cross-course connections helped this student have fun while reinforcing the interconnection between science disciplines. 
However, in a study of a youth after-school programming club, Fields and Enyedy found that the benefits of localized recognition did not automatically transfer to other settings~\cite{fields2013picking}. When a teacher expressly recognized students as experts, it was still important for students to be recognized by their peers as well. Despite differences in population, as Fields and Enyedy studied 11-year-olds and we focused on college students, we see a similar phenomenon and claim that what underpinned Jake's transformational experience with recognition was that it came from multiple sources, all of which reaffirmed the other. Designing new coursework activities, whether for youth or young adults, opens more avenues for students to be recognized for a renegotiated identity.

Engaging with code in the context of a computationally integrated physics course opened one such avenue for Jake's self-perception as a physicist, but the impact of recognition on Jake's PI proved complicated. Jake considered the recognition he received to be meaningful beyond the confines of the course and felt that it clarified his place in the field of physics more broadly, affirming him as a physicist in ways he had never been affirmed before. However, being recognized for his computational skills made Jake realize that he had not been recognized for his physics skills. A related theme also surfaced in a study of a computationally integrated introductory physics course in which students experienced a disconnect between their previously developed coding identities and this new context and when recognized for their skills by classmates, developed course-specific identities~\cite{bumler2019previous}. While the authors attributed this to students' perceptions of an inauthentic use of code, a similar occurrence surfaced in the multi-day activity on which this case study centered, which, we argue, did pose an authentic use of code by asking students to engage with advanced code written by someone else. The problem, moreover, was inconsistent code that disproportionately impacted Bridget, who had less computational resilience than Jake. Not only did the provided code have a distracting and confusing unused function, but it also differed from the style and use of code to which students had become acclimated. So, when encountering that inconsistency, Jake became motivated to rewrite it, and Bridget disengaged 
because that small inconsistency directly impacted the extent to which she felt she could understand the task at hand. Even though Bridget knew how to write nested \texttt{for} loops, she struggled with understanding someone else's code, as it was written differently than how she had been asked to write code this semester. Specifically, the built-in MATLAB functions confused Bridget because she had never seen them before and they weren't explained in the code. For Jake, this did not pose an issue, because he already had the latent skill set that this course had not leveraged thus far. Moments of frustration for Bridget served as moments of opportunity for Jake, turning the cognitive disparities due to differing amounts of computational exposure before class into affective disparities in engagement with the activity.


\section{Limitations}\label{Sec:limitations}
The theme explored in this paper surfaced during a more general study of how the integration of computation into a physics major course affected individual students' stances toward physics, computation, and their identity as a physics person. For the analysis presented here, we found an emergent theme that profoundly influenced the outlook of two students in opposite ways: they conceptualized and engaged with code differently, and as a result, developed self-perceptions with divergent affective tones. This theme surfaced during preliminary analysis and was investigated by representing students' narratives in conjecture maps, informed by observations of student behavior and interactions, and the products of this activity. Our conjecture maps do not communicate the entirety of Jake and Bridget's experiences with this multi-day activity, limited 
by our representative decisions as researchers. 
To mitigate this, we relied on a highly reflexive process that used analytic memos to elicit the most salient themes that surfaced both during data collection and while drafting broader narratives and conjecture maps of these students' experiences.

We find it important to note that our consideration of differing impacts only contrasted two students' experiences. 
However, we deliberately selected a small grain size for our case study so that we could meaningfully communicate the development of unique perspectives. 
We acknowledge that our study has limited generalizability, and we transparently presented the creation of our conjecture maps 
to provide rich descriptions that enable readers to determine resonance for their own contexts~\cite{tracy2010qualitative}.  

\section{Future Work}\label{Sec:future_work}
In this paper, we considered the development of students' physics identities in the context of a computationally integrated physics course. If, in these contexts, students change in the extent that they see themselves as physicists, the extent to which students see themselves as programmers can shift as well~\cite{bumler2019previous}. The divide between being a `physics person' and a `coding person' posed an affective tension for Jake, as he felt a contrast between being recognized for his coding knowledge rather than his physics knowledge. With Jake's experience in mind, we believe that the intersections of physics identity and coding identity are worthy of future investigation in computationally integrated physics courses and plan to pursue this line of inquiry.

At present, little is known about assessment of computation in a physics context~\cite{sabo2026we, burke2017developing}, not only in regards to practice best practices, but also considering impacts of grading schemes, both cognitive and affective. In this case study, Bridget's experience highlighted how grading mechanisms can serve as a factor that motivates engagement within a learning environment. In another affective case study of a computationally integrated physics course, ungraded code was seen as `busy work'~\cite{hamerski}. Taken together, these two experiences draw attention for a deeper understanding of the affective ramifications of grading code in computationally integrated physics courses. We look forward to seeing scholarship that addresses this gap in the literature so that instruction can foster cognitive and affective growth with research-based approaches to assessment.

\section{Conclusion}\label{Sec:conclusion}
Individual students experience encounters with computation differently, and these encounters shape students' attitudes towards themselves as physicists. To ensure that computation is a meaningful and affirming encounter for students, we need to understand how and why they develop different attitudes towards themselves as a result of their engagement.

In this paper, we presented a comparative case study of two students in a computationally integrated physics course who developed affectively divergent self-perceptions as a result of differing interpretations of and engagement with a multi-day activity. Our evidence predominately substantiates current understandings of epistemic framing and epistemic agency, and in so doing, draws attention to the affective impacts of activity design. Our modified use of conjecture mapping for narrative analysis helped clarify how these students' experiences reveal how self-perception can differ when computation is part of a physics student's educational experience, further complicating what it means to develop a physics identity. 

For practitioners, Bridget's experience highlights the potential for course-level consistency within the structure of assignments, even at the micro-level of syntax, to have affective impacts. Both courses and activities have an underlying ethos that students interpret and enact, and without explicit instruction, students activate their existing, and likely different, epistemic frames. 

We proposed this method as a suitable technique for narrative analysis and hope our analysis has demonstrated to fellow researchers how such a retrospective analysis of student experiences can connect to concrete activity design decisions. We see this case study as one of many steps to understand how students experience computation so that a body of research enables practitioners to employ diverse activity and curricular design, thereby empowering a wide variety of students to experience both cognitive and affective success when encountering computation in a physics context. 
Moreover, aligning instructional intent with impact creates the possibility to be hyper-responsive and 
revise instructional techniques for students so they can benefit from affirming experiences with computation.

\section{Acknowledgements}
This material is based upon work supported by the National Science Foundation under Grant No. 2236244 and 2237827. Any opinions, findings, and conclusions or recommendations expressed in this material are those of the author(s) and do not necessarily reflect the views of the National Science Foundation.

This work was also funded by a Norwegian Centennial Chair Travel Fellowship from the University of Minnesota (UMN). Thanks to Associate Professor Scharber of UMN's Department of Curriculum and Instruction for methodological expertise and to researchers in the Center for Computing in Science Education at the University of Oslo for fruitful exchanges of theoretical musings.

\bibliography{bibliography.bib} 

\end{document}